\documentclass[manuscript,noblind]{geophysics}

\usepackage[utf8x]{inputenc}
\usepackage{amsmath, amssymb, amsfonts, amsbsy, latexsym, amsthm}
\usepackage{algorithm, algorithmic,url}
\usepackage{verbatim, subfig}
\usepackage{hyperref}
\hypersetup{colorlinks,linkcolor={blue},citecolor={blue},urlcolor={red}}

\title{Wave simulation in non-smooth media by PINN with quadratic neural network and PML condition}
\date{\vspace{-5ex}}

\ms{} 

\address{
18b912018@stu.hit.edu.cn, aghamiry@geoazur.unice.fr,\\ operto@geoazur.unice.fr, jwm@pku.edu.cn\\
\footnotemark[1]Department of Mathematics and Center of Geophysics, Harbin Institute of Technology, Harbin 150001, China\\
\footnotemark[2]School of Earth and Space Science, Peking University, Beijing 100871, China\\
\footnotemark[3]Geoazur, Université Côte d’Azur, CNRS, IRD, OCA, Valbonne, France.
}

\author{Yanqi Wu\footnotemark[1]\footnotemark[3], Hossein S. Aghamiry\footnotemark[3], Stephane Operto\footnotemark[3], Jianwei Ma\footnotemark[1]\footnotemark[2]}

\begin{document}

\maketitle
\footer{}
\lefthead{}
\righthead{PINN for scattered wave equation}

\newpage

\begin{abstract}
Frequency-domain simulation of seismic waves plays an important role in seismic inversion, but it remains challenging in large models. The recently proposed physics-informed neural network (PINN), as an effective deep learning method, has achieved successful applications in solving a wide range of partial differential equations (PDEs), and there is still room for improvement on this front. For example, PINN can lead to inaccurate solutions when PDE coefficients are non-smooth and describe structurally-complex media. In this paper, we solve the acoustic and visco-acoustic scattered-field wave equation in the frequency domain with PINN instead of the wave equation to remove source singularity. We first illustrate that non-smooth velocity models lead to inaccurate wavefields when no boundary conditions are implemented in the loss function. Then, we add the perfectly matched layer (PML) conditions in the loss function of PINN and design a quadratic neural network to overcome the detrimental effects of non-smooth models in PINN. We show that PML and quadratic neurons improve the results as well as attenuation and discuss the reason for this improvement. We also illustrate that a network trained during a wavefield simulation can be used to pre-train the neural network of another wavefield simulation after PDE-coefficient alteration and improve the convergence speed accordingly. This pre-training strategy should find application in iterative full waveform inversion (FWI) and time-lag target-oriented imaging when the model perturbation between two consecutive iterations or two consecutive experiments can be small.
\end{abstract}
%
%
\section{Introduction}
The numerical simulation of seismic waves is an important topic in seismic exploration since its accuracy can impact wave-equation based imaging results \citep{Virieux_2009_TLE,Virieux_2011_RSP}. Moreover, it represents the most computationally-expensive ingredient of Full waveform inversion (FWI), which is a staple for high-resolution imaging of the Earth's interior in exploration geophysics and earthquake seismology \cite[]{Virieux_2017_FWI,Tromp_2019_SWI}. The wave equation can be solved either in time or frequency domains \cite[]{Plessix_2017_CAT,Kostin_2019_DFA}. Although explicit time-domain methods have low memory requirements and can be efficiently parallelized over sources, frequency-domain modeling is still attractive because efficient multiscale frequency-domain FWI on long-offset data can be implemented with a few discrete frequencies and attenuation effects of arbitrary complexity \cite[]{Virieux_2009_TLE}. However, frequency-domain modeling requires solving a large and sparse linear system per frequency, which is difficult to implement in large models, despite several methods based on direct and iterative solvers have been proposed to address this challenge \cite[]{Plessix_2007_HIS,Erlangga_2008_IMM,Amestoy_2016_FFF}.\\
Recently, deep learning (DL), combining machine learning and deep neural networks (NN), has made significant progress in solving problems with high-dimensional data, and it has proved to be effective at discovering intricate structures from huge, complex data sets \cite[]{Lecun_2015_DL}. Putting aside the idea of artificial intelligence (AI) and the successful applications in data science, DL has shown a bright prospect for solving partial differential equations (PDEs), especially in high-dimensional settings \cite[]{Burger_2021_CBD,Long_2018_PLP,Sirignano_2018_DGM,Berg_2019_DDP}. Generally, DL-based PDE solving methods can be divided into two broad categories \cite[]{Karniadakis_2021_PIM}: 
\begin{itemize}
\item Minimizing the difference between network prediction and real data when there are a lot of data and ambiguous physical rules \cite[]{Han_2018_SHP,Long_2019_PLP}, 
\item Minimizing the governing-equation errors when there are few data (boundary and initial conditions) and precisely known physical rules \cite[]{Zhu_2019_PDL,Sun_2020_SMF}. 
\end{itemize}
Physics-informed neural networks (PINNs), seamlessly integrating both the data-based and mathematical model-based terms, has flexibility in informing physical laws described in differential equations and thus achieving successful applications in solving a wide range of PDEs \cite[]{Cai_2022_PNN,Pang_2019_FFP,Fang_2019_PNN,Zhang_2020_LMS}.
PINNs output the PDE solutions by a NN that is trained to minimize the PDE residuals on the part of points in the certain domain \cite[]{Raissi_2019_PNN}. They use automatic differentiation to calculate the partial derivatives in PDEs without the explicit need for mesh generation. Both the forward and inversion problems could be solved in this framework. Since it is discretization-free, PINNs can deal with PDEs in high-dimensional domains, which is difficult in conventional numerical methods. \\
However, PINN methods can not defeat the traditional numerical methods yet because they are not accurate or efficient enough for solving standard forward problems. PINN suffers from a slow convergence rate and accuracy degradation for a certain class of PDEs since it recasts the forward problem as the minimization of a loss function for estimating the weights of the NN. Then, the efficiency is limited by the convergence rate during network training because of the high-dimensional non-convex loss function and the limited accuracy induced by the size of the network. Based on the universal approximation theorem, the objective solution should be smooth when a NN is limited in size \cite[]{Barron_1993_UAB}.\\
Even with those limitations, attracted by the potential of PINN, many researchers have been implementing the method for improvement. 
A first approach aims at optimizing the NN architecture:  Using adaptive activation functions, \cite{Jagtap_2020_AAF} significantly improved the convergence rate, especially at early training, as well as the solution accuracy of PINN. \cite{Gao_2021_PPG} proposed a novel physics-constrained CNN learning architecture to learn solutions of parametric PDEs on irregular domains without any labeled data. 
To deal with PDEs with a non-smooth solution, \cite{Kharazmi_2021_HVP} proposed a variational formulation of PINN based on the Galerkin method (hp-VPINN) and \cite{Mao_2020_PNN} studied how the chosen positions of the training samples impact the performances for smooth solutions and Riemann problems.\\
For wave propagation problems, most of the research focus on time domain wave equation. Both of the forward and inversion problems are studied with PINN. 
\cite{Xu_2019_PIN} performed the velocity inversion with PINN in simple inhomogenous media and obtained a better result than FWI. \cite{Voytan_2020_WPP} extended PINNs to inhomogeneous velocity model and implementing absorbing boundary conditions. \cite{Karimpouli_2020_PIM} solved a 1D form seismic wave equation and simultaneously invert the velocity model. \cite{Moseley_2020_SWE} proved that PINN can model and predict a wide range of physical phenomena by only given the first few timesteps of the solution. \cite{Huang_2021_SPD}  eliminated the source singularity and used a multi-scale deep neural network with periodic activation function to improve the accuracy and convergence speed of the PINNs method. \cite{Rasht_2022_PNN} applied PINNs to seismic inversions within heterogeneous media and demonstrated that PINNs yield excellent results for inversions on all cases considered and with limited computational complexity. 
For the wave propagation in frequency domain, \cite{Song_2021_SFA} and \cite{Song_2022_WRI} used PINN to solve the scattered-field (Lippmann-Schwinger) wave equation instead of the total-field wave equation to remove the singularity and complexity of the wavefield near the source. However, sharp contrasts in the PDE coefficients (in this case, velocity contrasts in the underground medium) make the scattered wavefield more complex and non-smooth and hence affect the accuracy of PINN. In this paper, we address the problem of non-smooth velocity models with two recipes. First, we add absorbing perfectly matched layer (PML) conditions \citep{Berenger_1994_PML} into PINN to explicitly strengthen the coupling of the two outputs of PINN (in this case, the real and imaginary part of the wavefield) on the boundaries. 
Second, we use quadratic neurons instead of the traditional linear neurons in the first layer of the NN to increase its nonlinearity and improve the estimation of the complex scattered wavefield. We show the improved wavefield reconstruction induced by these two ingredients with the Marmousi model.
We also solve the scattered-field wave equation in a complex-valued velocity model, such that the two outputs of PINN could be explicitly coupled everywhere in the medium.
We first illustrate how non-smooth velocity models deteriorate the accuracy of the wavefields computed with PINN. Then, we show and discuss why PMLs mitigate this issue and contribute to reconstructing more accurate wavefields and why PINN behaves better in a complex-valued velocity model. We also propose a pre-training workflow that mitigates the computational burden of wavefield simulation after model alteration. This pre-training strategy can find interesting applications in iterative FWI and time-lapse imaging to mitigate the computational burden of the forward problem in iterations or during the monitoring stage.

\section{Theory}
The non-smooth behavior of monochromatic wavefields near the point source makes the wave equation challenging to solve with PINN, especially when the network is not excessively large. 
To bypass this issue, \cite{Alkhalifah_2021_WSM} proposed to solve the scattered-field wave equation (Lippmann-Schwinger equation) to avoid the singularity at source. In the frequency-domain, this equation is given by
\begin{equation}\label{1}
\begin{aligned}
\left(\omega^{2} m[\mathbf{x}] +\nabla^{2}\right) \delta {u}[\mathbf{x}] =-\omega^{2} \left(m[\mathbf{x}]-m_{0}[\mathbf{x}]\right) u_{0}[\mathbf{x}],
\end{aligned}
\end{equation}
\noindent where $\omega$ is the angular frequency, $m$ is the squared-slowness model as a function of space coordinates $\mathbf{x}=(x,z)$, $\nabla^{2}$ is the Laplace operator, $\delta u$ is the complex-valued scattered wavefield, $m_{0}$ is the background squared-slowness model and $u_{0}$ is the background wavefield in $m_0$. Moreover, $\delta u=\delta u_{r}+\hat{i}\delta u_{i}$, where $\delta u_{r}$ and $\delta u_{i}$ are the real and imaginary parts of the scattered wavefield and $\hat{i}=\sqrt{-1}$. In this study, we use a homogeneous background model, leading to
\begin{equation}\label{2}
\begin{aligned}
u_{0}[\mathbf{x}]=\frac{\hat{i}}{4} H_{0}^{(2)}\left({\sqrt{m}_0\omega}\left|\mathbf{x}-\mathbf{x}_{s}\right|\right),
\end{aligned}
\end{equation}
\noindent where $H_{0}^{(2)}$ is the zero-order Hankel function of the second kind, $m_{0}$ is the constant squared slowness and $\mathbf{x}_{s}$ denotes the position of the source \citep{Carcione_2015_WFR}.\\
Absorbing boundary conditions are classically used to simulate wave propagation in infinite media with classical numerical methods. The scattered-field wave equation with perfectly matched layer (PML) absorbing boundary conditions \citep{Berenger_1994_PML} reads
\begin{equation}\label{4}
\begin{aligned}
&\frac{\partial}{\partial {x}}\left( \frac{e_{z}[z]}{e_{x}[x]} \frac{\partial \delta u[\mathbf{x}]}{\partial {x}}\right)+\frac{\partial}{\partial {z}}\left( \frac{e_{x}[x]}{e_{z}[z]} \frac{\partial \delta u[\mathbf{x}]}{\partial {z}}\right)
+\omega^{2} e_{x}[x] e_{z}[z] ~ m[\mathbf{x}]\delta u[\mathbf{x}] = \\
&-\omega^{2} e_{x}[x] e_{z}[z] ~  \left({m}[\mathbf{x}]-{m}_{0}[\mathbf{x}]\right) u_{0}[\mathbf{x}],
\end{aligned}
\end{equation}
\noindent where $e_{x}[x]=1-\hat{i} \frac{\sigma_{x}[x]}{\omega}$ and $e_{z}[z]=1-\hat{i}\frac{\sigma_{z}[z]}{\omega}$ \citep{Chen_2013_OFD}. The damping function $\sigma_{x}[x]$ (and similarly for $\sigma_{z}[z]$) defined as
\begin{equation}\label{5}
\begin{aligned}
\begin{gathered}
\sigma_{x}[x]= \begin{cases}2 \pi a_{0} f_{0}\left(\frac{l_{x}[x]}{L_{PML}}\right)^{2}, & \text { inside PML } \\ 0, & \text { outside PML }\end{cases},
\end{gathered}
\end{aligned}
\end{equation}
\noindent where $f_{0}$ is the peak frequency of the source, $L_{PML}$ is PML thickness, and $l_{x}[x]$ is the horizontal distance from the point $(x, z)$ in the PML to the PML-medium interface \citep{Chen_2013_OFD}. Moreover, $a_{0}$ is a constant set to 1.79 \cite[]{Zeng_2001_APM}. 
 
\subsection{PINN for the scattered wave equation with PML}
Frequency-domain seismic wave modeling (Helmholtz problem) is a boundary-value problem. It is classically tackled by solving a large and sparse indefinite linear system per frequency with direct or iterative methods, which is a challenging task for large-scale problems \citep{Ernst_2011_Helm,Dolean_2015_IDD}. The PINN method, which is mesh-free, bypasses this linear-algebra problem by minimizing the following loss function estimated at $N$ random spatial positions \cite[]{Alkhalifah_2021_WSM}:
\begin{equation}\label{3}
\begin{aligned}
f[\theta] &= \frac{1}{N} \sum_{j=1}^{N} \left|\omega^{2} m^{(j)} \delta u_{r}^{(j)}[\theta]+\nabla^{2} \delta u_{r}^{(j)}[\theta]+\omega^{2}\left(m-m_{0}\right)^{(j)} u_{0 r}^{(j)}\right|^{2}\\
&+\frac{1}{N} \sum_{j=1}^{N}\left|\omega^{2} m^{(j)} \delta u_{i}^{(j)}[\theta]+ \nabla^{2} \delta u_{i}^{(j)}[\theta]+\omega^{2}\left(m-m_{0}\right)^{(j)} u_{0 i}^{(j)}\right|^{2},
\end{aligned}
\end{equation}
where $\bullet^{(j)}$ denotes the training sample index.
The inputs to the network, like a function, are $N$ spatial locations $\bf{x}$, where $N$ refers to the number of training samples. The output of the network are the real and imaginary values of the scattered wavefield ($\delta u_{r}$ and $\delta u_{i}$, respectively), which are functions of the optimization variables $\theta$, namely the weights of the NN.
We decompose the loss function into two terms related to the real and imaginary parts of the governing-equation errors because NN cannot manage complex numbers for now. It is worth noting that the first and the second terms involve only $\delta u_{r}$ and $\delta u_{i}$, respectively, when $m$ is real valued and no boundary conditions are explicitly implemented in PINN. This suggests that the loss function of equation \ref{3} may lack some physical constraints considering that, the wavefields being complex valued, $\delta u_{r}$ and $\delta u_{i}$ are closely related.
Indeed, \citet{Alkhalifah_2021_WSM} and \citet{Rasht_2021_PNN} showed that PINN can theoretically simulate waves in infinite media from the Helmholtz equation without boundary conditions. The reason we can respectively deal with $\delta u_{r}$ and $\delta u_{i}$ in equation \ref{3} without boundary conditions is that their relationship is implicitly informed by $u_{0}$ and subsurface properties. 
Here, we propose however to enforce this relationship more explicitly in PINN by implementing the scattered wave equation with PML, equation~\ref{4}, in the loss function.
%
The loss function $f[\theta]$ with PML reads
\begin{equation}\label{9}
\begin{aligned}
f_{PML}[\theta]&=\frac{1}{N} \sum_{j=1}^{N}\left(F_{r}\left[ \delta u_r^{(j)}[\theta], \delta u_i^{(j)}[\theta]\right]^{2} +F_{i}\left[\delta u_r^{(j)}[\theta], \delta u_i^{(j)}[\theta]\right]^{2}\right),
\end{aligned}
\end{equation}
where the real and imaginary parts of the equation errors are
\begin{equation}\label{7}
\begin{gathered}
\begin{aligned}
F_r[\delta u_r, \delta u_i]=&\frac{\partial}{\partial {x}}\left(\frac{1+c^{2} l_{x}^{2} l_{z}^{2}}{1+c^{2} l_{x}^{4}} \frac{\partial \delta u_{r}}{\partial {x}}+\frac{c\left(l_{z}^{2}-l_{x}^{2}\right)}{1+c^{2} l_{x}^{4} } \frac{\partial \delta u_{i}}{\partial {x}}\right)\\
&+\frac{\partial}{\partial z}\left(\frac{1+c^{2} l_{x}^{2} l_{z}^{2}}{1+c^{2} l_{z}^{4}} \frac{\partial \delta u_{r}}{\partial {z}}+\frac{c\left(l_{x}^{2}-l_{z}^{2}\right)}{1+c^{2} l_{z}^{4} } \frac{\partial \delta u_{i}}{\partial {z}}\right) \\
&+\left(1-c^{2} l_{x}^{2} l_{z}^{2}\right) \omega^{2} \left( m \delta u_{r} + \left(m-m_{0}\right) u_{0r} \right)\\
&+c\left(l_{x}^{2}+l_{z}^{2}\right) \omega^{2} \left( m \delta u_{i}+ \left(m-m_{0}\right) u_{0i} \right),
\end{aligned}
\end{gathered}
\end{equation}
and
\begin{equation}\label{8}
\begin{gathered}
\begin{aligned}
\begin{aligned}
F_i[\delta u_r, \delta u_i]=&\frac{\partial}{\partial x}\left(\frac{c\left(l_{x}^{2}-l_{z}^{2}\right)}{1+c^{2} l_{x}^{4}} \frac{\partial \delta u_r}{\partial x}+\frac{1+c^{2} l_{x}^{2} l_{z}^{2}}{1+c^{2} l_{x}^{4}} \frac{\partial \delta u_i}{\partial x}\right)\\
&+\frac{\partial}{\partial z}\left(\frac{c\left(l_{z}^{2}-l_{x}^{2}\right)}{1+c^{2} l_{z}^{4}} \frac{\partial\delta u_{r}}{\partial z}+\frac{1+c^{2} l_{x}^{2} l_{z}^{2}}{1+c^{2} l_{z}^{4}} \frac{\partial \delta u_i}{\partial z}\right) \\
&-c\left(l_{x}^{2}+l_{z}^{2}\right) \omega^{2}\left( m \delta u_{r} + \left(m-m_{0}\right) u_{0r} \right)\\
&+\left(1-c^{2} l_{x}^{2} l_{z}^{2}\right) \omega^{2} \left( m \delta u_{i}+ \left(m-m_{0}\right) u_{0i} \right),
\end{aligned}
\end{aligned}
\end{gathered}
\end{equation}
$c=\frac{2 \pi a_{0} f_{0}}{\omega L_{P M L}^{2}}$ and $u_{0r}$, $u_{0i}$ are computed with the closed-form expression given in equation~\ref{2}. One can readily check that the two terms of the loss function $f_{PML}$ are now functions of both $\delta u_{r}$ and $\delta u_{i}$ unlike those of equation~\ref{3}. \\
Aside from all the constant coefficients in equation \ref{5}, we also need to compute during the training the derivative of $l_{x}[x]$ and $l_{z}[z]$ without relying on mesh. This is performed in $tensorflow$ by {\it{Relu}} activation function:
\begin{equation}\label{6}
\begin{aligned}
l_{x}[x]=&\max \left(0,x_{b 1}-x\right)+\max \left(0,x-x_{b 2}\right)\\
=&\operatorname{Relu}\left(x_{b 1}-x\right)+\operatorname{Relu}\left(x-x_{b 2}\right),
\end{aligned}
\end{equation}
where $x_{b1}$ and $x_{b2}$ denote the position of the boundaries. The functions $\sigma_{z}$ and $l_{z}$ can be chosen similarly. \\
Although PMLs explicitly provide some physical constraints, like the relationship between the real and imaginary parts of the wavefield, it only works on the boundary. In attenuating media, the velocity model is complex-valued, which can couple the real and imaginary parts of wavefield everywhere in space. Moreover, attenuating effects are also necessary to represent the real physics and can be part of optimization variables during the imaging inverse problem \citep[e.g.,][]{Operto_2018_MFF}.
Here, we implement attenuation with the Kolsky-Futterman (KF) model. In this case, the squared slowness reads
\begin{equation}\label{KF}
\begin{gathered}
\begin{aligned}
\begin{aligned}
m[\mathbf{x}]=\frac{1}{v^{2}[\mathbf{x}]}\left(1-\frac{\alpha[\mathbf{x}]}{\pi} \ln \left|\frac{\omega}{\omega_{r}}\right|+\hat{i} \frac{\alpha[\mathbf{x}]}{2}\right)^{2},
\end{aligned}
\end{aligned}
\end{gathered}
\end{equation}
where $\omega_{r}$ is a reference angular frequency, $v$ and $\alpha$ are the real-valued phase velocity and attenuation factor (inverse of quality factor $Q$) at the reference frequency
The complex-squared slowness can be decomposed to the real ($m_{r}$) and imaginary ($m_{i}$) parts as
\begin{equation}\label{complexm}
\begin{gathered}
\begin{aligned}
\begin{aligned}
m[\mathbf{x}] =& m_{r}[\mathbf{x}]+\hat{i} m_{i}[\mathbf{x}], 
\end{aligned}
\end{aligned}
\end{gathered}
\end{equation}
where  
\begin{equation}\label{complexm1}
\begin{gathered}
\begin{aligned}
\begin{aligned}
m_{r}[\mathbf{x}] =& \frac{1}{v^{2}[\mathbf{x}]}\left(1-\frac{\alpha[\mathbf{x}]}{\pi} \ln \left|\frac{\omega}{\omega_{r}}\right|\right)^{2}-\frac{\alpha^{2}[\mathbf{x}]}{4 v^{2}[\mathbf{x}]}, \\
m_{i}[\mathbf{x}] =& \frac{\alpha[\mathbf{x}]}{v^{2}[\mathbf{x}]}\left(1-\frac{\alpha[\mathbf{x}]}{\pi} \ln \left|\frac{\omega}{\omega_{r}}\right|\right).
\end{aligned}
\end{aligned}
\end{gathered}
\end{equation}
Then, the real and imaginary parts of the equation errors with complex-valued slowness are:
\begin{equation}\label{mrpinn}
\begin{gathered}
\begin{aligned}
F_r[\delta u_r, \delta u_i]=&\frac{\partial}{\partial {x}}\left(\frac{1+c^{2} l_{x}^{2} l_{z}^{2}}{1+c^{2} l_{x}^{4}} \frac{\partial \delta u_{r}}{\partial {x}}+\frac{c\left(l_{z}^{2}-l_{x}^{2}\right)}{1+c^{2} l_{x}^{4} } \frac{\partial \delta u_{i}}{\partial {x}}\right)\\
&+\frac{\partial}{\partial z}\left(\frac{1+c^{2} l_{x}^{2} l_{z}^{2}}{1+c^{2} l_{z}^{4}} \frac{\partial \delta u_{r}}{\partial {z}}+\frac{c\left(l_{x}^{2}-l_{z}^{2}\right)}{1+c^{2} l_{z}^{4} } \frac{\partial \delta u_{i}}{\partial {z}}\right) \\
&+\left(1-c^{2} l_{x}^{2} l_{z}^{2}\right) \omega^{2} \left( m_{r} \delta u_{r} - m_{i}\delta u_{i} + \left( m_{r} -m_{0r}\right) u_{0r} - \left( m_{i}-m_{0i} \right)u_{0i} \right)\\
&+c\left(l_{x}^{2}+l_{z}^{2}\right) \omega^{2} \left( m_{r} \delta u_{i} + m_{i} \delta u_{r} + \left(m_{r}-m_{0r}\right) u_{0i} + \left(m_{i}-m_{0i}\right) u_{0r} \right),
\end{aligned}
\end{gathered}
\end{equation}
and
\begin{equation}\label{mipinn}
\begin{gathered}
\begin{aligned}
\begin{aligned}
F_i[\delta u_r, \delta u_i]=&\frac{\partial}{\partial x}\left(\frac{c\left(l_{x}^{2}-l_{z}^{2}\right)}{1+c^{2} l_{x}^{4}} \frac{\partial \delta u_r}{\partial x}+\frac{1+c^{2} l_{x}^{2} l_{z}^{2}}{1+c^{2} l_{x}^{4}} \frac{\partial \delta u_i}{\partial x}\right)\\
&+\frac{\partial}{\partial z}\left(\frac{c\left(l_{z}^{2}-l_{x}^{2}\right)}{1+c^{2} l_{z}^{4}} \frac{\partial\delta u_{r}}{\partial z}+\frac{1+c^{2} l_{x}^{2} l_{z}^{2}}{1+c^{2} l_{z}^{4}} \frac{\partial \delta u_i}{\partial z}\right) \\
&-c\left(l_{x}^{2}+l_{z}^{2}\right) \omega^{2} \left( m_{r} \delta u_{r} - m_{i}\delta u_{i} + \left( m_{r} -m_{0r}\right) u_{0r} - \left( m_{i}-m_{0i} \right)u_{0i} \right)\\
&+\left(1-c^{2} l_{x}^{2} l_{z}^{2}\right) \omega^{2} \left( m_{r} \delta u_{i} + m_{i} \delta u_{r} + \left(m_{r}-m_{0r}\right) u_{0i} + \left(m_{i}-m_{0i}\right) u_{0r} \right),
\end{aligned}
\end{aligned}
\end{gathered}
\end{equation}
where $m_{0r}$ is the real part and $m_{0i}$ is the imaginary part of the background squared-slowness model.

\subsection{Quadratic neural network}
Although replacing the full wavefield with the scattered wavefield can address the source singularity problem, the scattered wavefield still contains the rough components of the total field. This may be one of the reasons why PINN reconstructed a smoother solution than the exact one \cite[]{Alkhalifah_2021_WSM}. We designed the NN with eight hidden layers with 80 neurons in every layer. The activation function between layers, other than the last layer, is a hyperbolic tangent in all the examples. The last hidden layer connected to the output layer is linear. Adam optimizer is chosen to train the NN. Increasing the size of the NN should lead to a more accurate reconstruction, albeit with an increased computational cost. 
Here, we propose using quadratic neurons instead of the traditional linear neuron at the first layer of NN to increase the nonlinearity of the NN and improve the estimation of the scattered wavefield accordingly. Let us take a look at one neuron of the deep NN. Usually, the output $x^{(1)}$ of a neuron in the first layer is given by:
\begin{equation}\label{10}
\begin{gathered}
\begin{aligned}
x^{(1)}=\tanh \left(x w_{1}+z w_{2}+b\right),
\end{aligned}
\end{gathered}
\end{equation}
\noindent where $w_{1}$, $w_{2}$ and $b$ are parameters to be optimized, the activation function is hyperbolic tangent and $x$ and $z$ are the inputs of the network, in this case, the space coordinates of the wavefield. As shown in Figure \ref{fig:net}, the improved formulation is given by:
\begin{equation}\label{11}
\begin{gathered}
\begin{aligned}
\begin{aligned}
x^{(1)}=\tanh \left(x w_{1}+z w_{2}+x z w_{3}+x^{2} w_{4}+z^{2} w_{5}+b\right).
\end{aligned}
\end{aligned}
\end{gathered}
\end{equation}
From the NN perspective, it increases the non-linearity and flexibility by only adding a few parameters leading to a negligible computational overhead. From a physical standpoint, wavefield is closely related to the distance from the current point to the source, which is easier to describe with the quadratic of space coordinate components. 
%
%
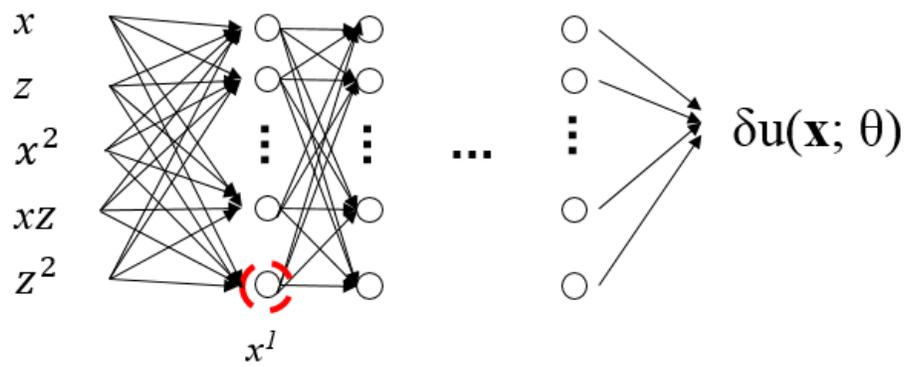
\plot{net}{width=0.8\textwidth}{Illustration of quadratic neural network.}
%
%
%
\section{Numerical modeling}
\renewcommand{\figdir}{Fig}
We first solve the frequency-domain scattered-field wave equation in a smoothed Marmousi model for $f=5 Hz$ frequency  (Figure~\ref{fig:fig2}a). We assess the accuracy of the wavefield reconstructed by PINN against the wavefield computed with a finite-difference frequency-domain (FDFD) method  with adaptive FD coefficients \cite[]{Chen_2013_OFD}. Five thousand training points are randomly chosen from a 100 $\times$ 100 Cartesian grid in every epoch. 
The grid intervals in $x$ and $z$ are $40 m$ and $20 m$, respectively. The source is at the middle of the surface. 
It is reminded that the NN is formed by eight hidden layers with 80 neurons in every layer. The activation function between layers, other than the last layer, is an hyperbolic tangent in all the examples. The last hidden layer connected to the output layer is linear. 
We train the network with Adam optimizer and 150000 epochs on NVIDIA Tesla V100 32 GB GPU. The scattered wavefield computed with PINN when the loss function doesn't contain PML condition are shown in Figures~\ref{fig:fig2}b-\ref{fig:fig2}d. These results suggest that PINN manages to solve the scattered-field wave equation in the smoothed velocity model with acceptable accuracy, although some mild differences with the FDFD wavefield are shown. However, if we replace the smoothed Marmousi model by the original Marmousi model (Figure~\ref{fig:fig3}a), the wavefield reconstructed by PINN is highly inaccurate (Figures~\ref{fig:fig3}c and \ref{fig:fig3}d). Moreover, this inaccurate reconstruction is barely improved when increasing the size of the NN. In fact, we have tested several NNs, e.g., 6 and 10 hidden layers (depth) with 80 neurons in every layer, eight hidden layers with 80 and 60 neurons in every layer, or a network given by {128, 128, 64, 64, 32, 32, 16, 16, 8, 8} neurons in every hidden layer from input to output without significant improvements relative to the results of Figures~\ref{fig:fig2}c and \ref{fig:fig2}d.
%
%
%
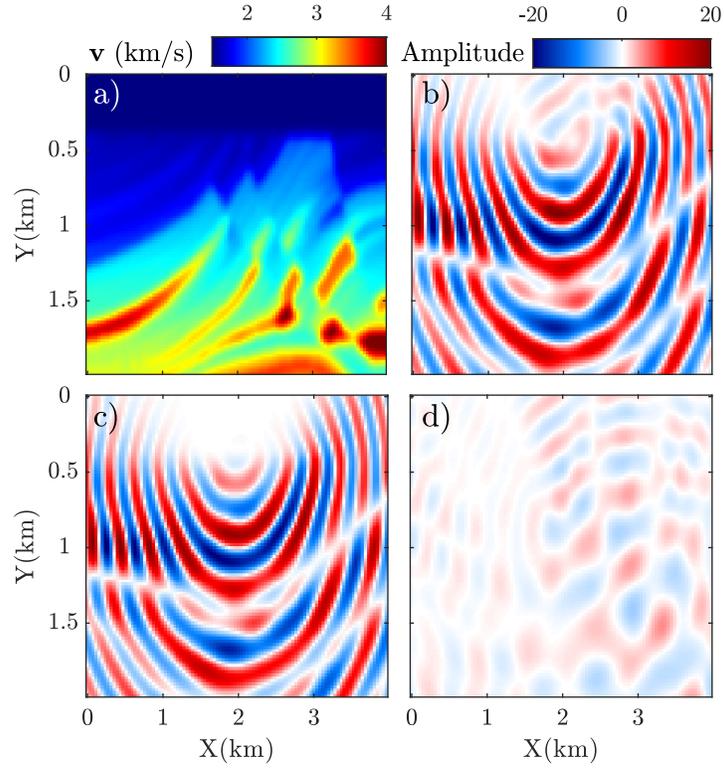
\plot{fig2}{width=0.62\textwidth}{The simulation results in smoothed Marmousi model. (a) Smoothed Marmousi model. (b) FDFD scattered fields (real part). (c) PINN scattered fields (real part). (d) Difference between Figure \ref{fig:fig2}b and Figure \ref{fig:fig2}c.}
%
%
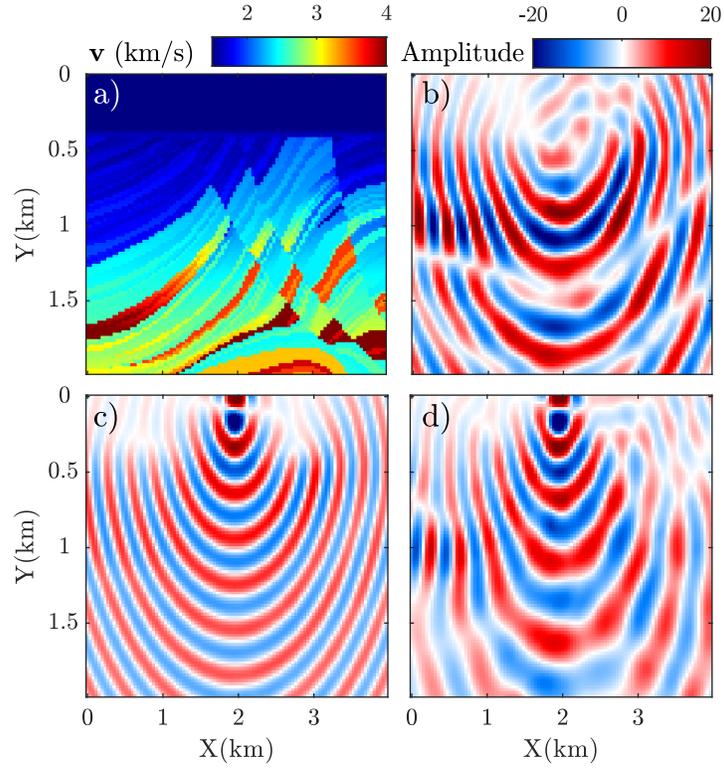
\plot{fig3}{width=0.62\textwidth}{The simulation results in non-smoothed Marmousi model. (a) Non-smoothed Marmousi model. (b) FDFD scattered fields (real part). (c) PINN scattered fields (real part). (d) Difference between Figure \ref{fig:fig3}b and Figure \ref{fig:fig3}c. }
%
%
\subsection{PINN with PML}
Since the network design barely changed the solution, a reasonable statement is that inaccurate wavefield reconstruction not only results from the network but also from the loss function. 
Considering equation \ref{3}, we notice that the real and imaginary parts of the scattered wavefield are calculated to some extent separately by minimization of the two terms of the loss function while they are closely physically entwined. In the FDFD method, the wavefield is obtained by computing the inverse of a large and sparse matrix $\mathbf{A}$ with PML absorbing conditions. Thus, the real and imaginary parts of the wavefield are both closely dependent on the structure of $\mathbf{A}$ and the boundary conditions. However, in equation \ref{3}, $\delta u_{r}[\theta]$ and $\delta u_{i}[\theta]$ depend on the parameters of the NN, and the relationship between the real and imaginary part is only implicitly informed to the NN through the background wavefield $u_{0 r}$ and $u_{0 i}$ in the right-hand side of the scattered field equation. Moreover, the non-smoothness of the velocity model will also increase the difficulty of learning the coupled relationship between the real and imaginary parts of the scattered field since the sharp contrasts of the medium are transferred in the scattering (or contrast) source. 
Accordingly, we repeat the simulation in the non-smooth Marmousi model using the loss function with PML, equation~\ref{9}, to explicitly enforce physical constraints on the boundaries and couple the real and imaginary parts of the scattered wavefield in the two terms of the loss function. Although PINN may automatically satisfy the wave equation in infinite media without implementing boundary conditions, a PML condition should not harm it. More insights about the effects of the PML during the training performed by PINN are illustrated in the final Discussion section.

The reconstructed wavefield (Figures~\ref{fig:fig4}a and \ref{fig:fig4}b)  shows how adding PML condition in the loss function significantly improves the wavefield reconstruction compared to Figures~\ref{fig:fig3}c and \ref{fig:fig3}d. However, compared to the ground truth shown in \ref{fig:fig3}b, the wavefield in Figure \ref{fig:fig4}a remains not accurate enough.
%
%
%
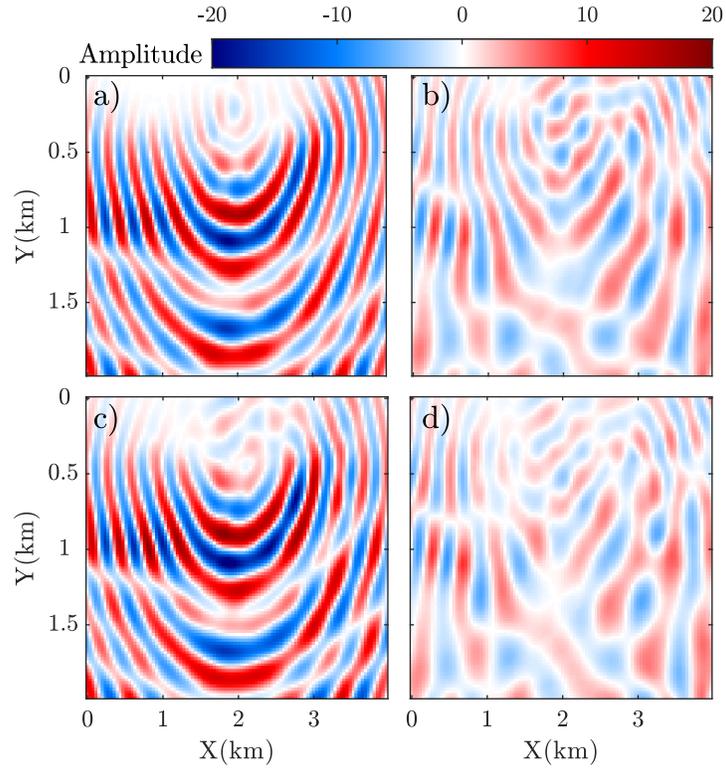
\plot{fig4}{width=0.62\textwidth}{The simulation results in non-smoothed Marmousi model with PML and quadratic NN. (a) PINN scattered field with PML (real part). (b) Difference between Figure \ref{fig:fig3}b and Figure \ref{fig:fig4}a. (c) PINN scattered field with PML and quadratic NN (real part). (d) Difference between Figure \ref{fig:fig3}b and Figure \ref{fig:fig4}c.}
%
%
%
%
\subsection{PINN with PML and quadratic neural network}
According to the wave equation, a non-smooth velocity model will cause a non-smooth (or discontinuous) Laplacian of the wavefield. However, the conventional NN with hyperbolic tangent as activation function is differentiable in any order. We may improve the NN by increasing its size, but this will generate significant computational overhead.
Instead, we propose to use quadratic neurons instead of linear neurons on the first layer. Figures~\ref{fig:fig4}c and \ref{fig:fig4}d show the wavefield reconstructed by PINN + PML + quadratic NN. Compared to PINN + PML shown in Figures~\ref{fig:fig4}a and \ref{fig:fig4}b, the wavefield reconstruction is improved with quadratic NN, especially near the surface. These more accurate wavefield reconstructions are indeed critical for future FWI applications.

\subsection{PINN in attenuating medium}
Implementing intrinsic attenuation in the underground medium makes the squared-slowness model complex-valued. In addition to better representing the physics of wave propagation into the Earth, attenuation couples the real and imaginary part of the wavefield at every spatial position in each term of the loss function, equations~\ref{mrpinn} and \ref{mipinn}, hence strengthening physical constraints in PINN. We apply attenuation with the homogeneous quality factor of $Q=15$ and repeat the wavefield reconstruction using PINN (Figures \ref{fig:fig6}c), PINN + PML (Figures \ref{fig:fig7}a) and PINN + PML + quadratic NN (Figures \ref{fig:fig7}c). After considering the attenuation, the wavefield reconstructed by PINN (Figure \ref{fig:fig6}c) is still inaccurate, although it is slightly improved compared to the wavefields (Figure \ref{fig:fig3}c) computed in the non-attenuating medium. Moreover, the visco-acoustic wavefields computed with PINN + PML (Figure \ref{fig:fig7}a) and PINN + PML + quadratic NN (Figure \ref{fig:fig7}c) contain more details than the corresponding wavefields computed without attenuation (Figure \ref{fig:fig4}a and \ref{fig:fig4}c) although the diffusive effects of attenuation, hence suggesting a more accurate wavefield reconstruction, especially in the shallow area.
%
%
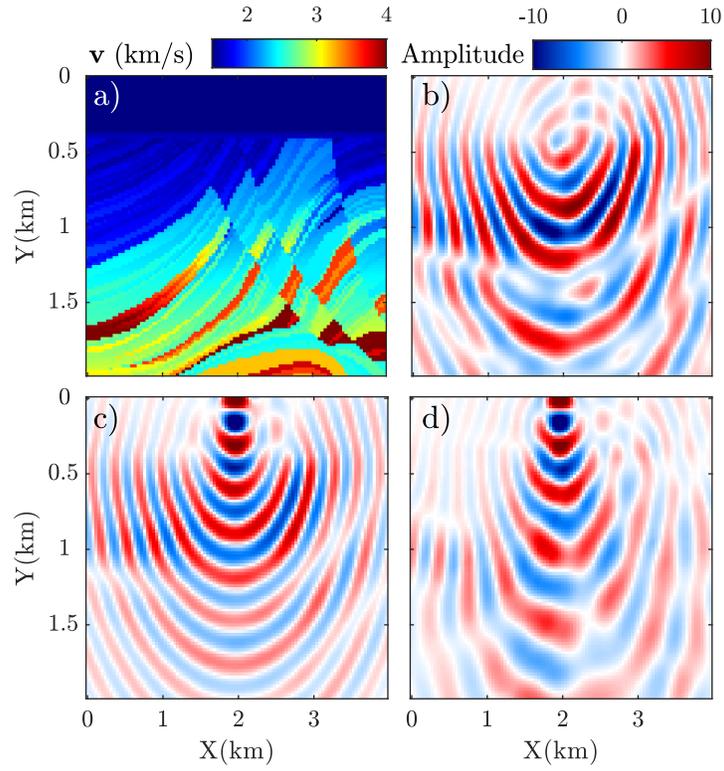
\plot{fig6}{width=0.62\textwidth}{The simulation results in attenuating Marmousi model with homogeneous quality factor Q=15. (a) Non-smoothed Marmousi model. (b) FDFD scattered fields (real part). (c) PINN scattered fields (real part). (d) Difference between Figure \ref{fig:fig6}b and Figure \ref{fig:fig6}c.}
%
%
%
%
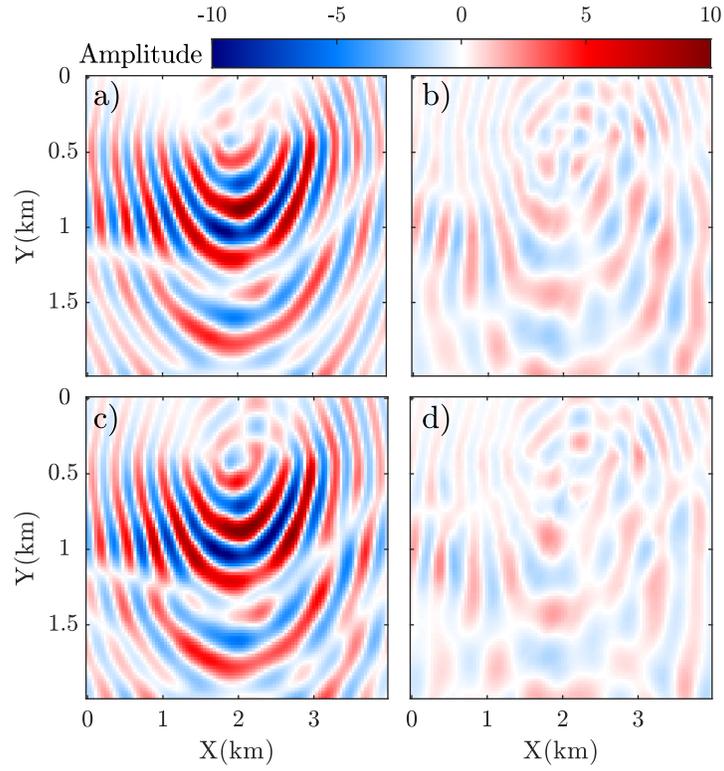
\plot{fig7}{width=0.62\textwidth}{The simulation results in attenuating Marmousi model with PML, quadratic NN and Q=15. (a) PINN scattered field with PML (real part). (b) Difference between Figure \ref{fig:fig6}b and Figure \ref{fig:fig7}a. (c) PINN scattered field with PML and quadratic NN (real part). (d) Difference between Figure \ref{fig:fig6}b and Figure \ref{fig:fig7}c.}
%
%
%
%
To relate more quantitatively the impact of modeling errors on velocity reconstruction, we take advantage of the bi-linearity of the wave equation \cite[]{Aghamiry_2019_IWR} to reconstruct the velocity model, as shown in Figures~\ref{fig:fig8}, from the simulated monochromatic wavefields. The bilinearity of the wave equation implies that the squared slowness model can be inferred exactly from a monochromatic wavefield by the following pointwise division:
\begin{equation}
m[\mathbf{x}]= \frac{b[\mathbf{x},\omega] - \nabla^2 u[\mathbf{x},\omega]}{\omega^2 u[\mathbf{x},\omega]},
\label{eqbili}
\end{equation}
where $b[\mathbf{x},\omega]$ is the physical source of the Helmholtz (total-field) equation. \\
%
%
%
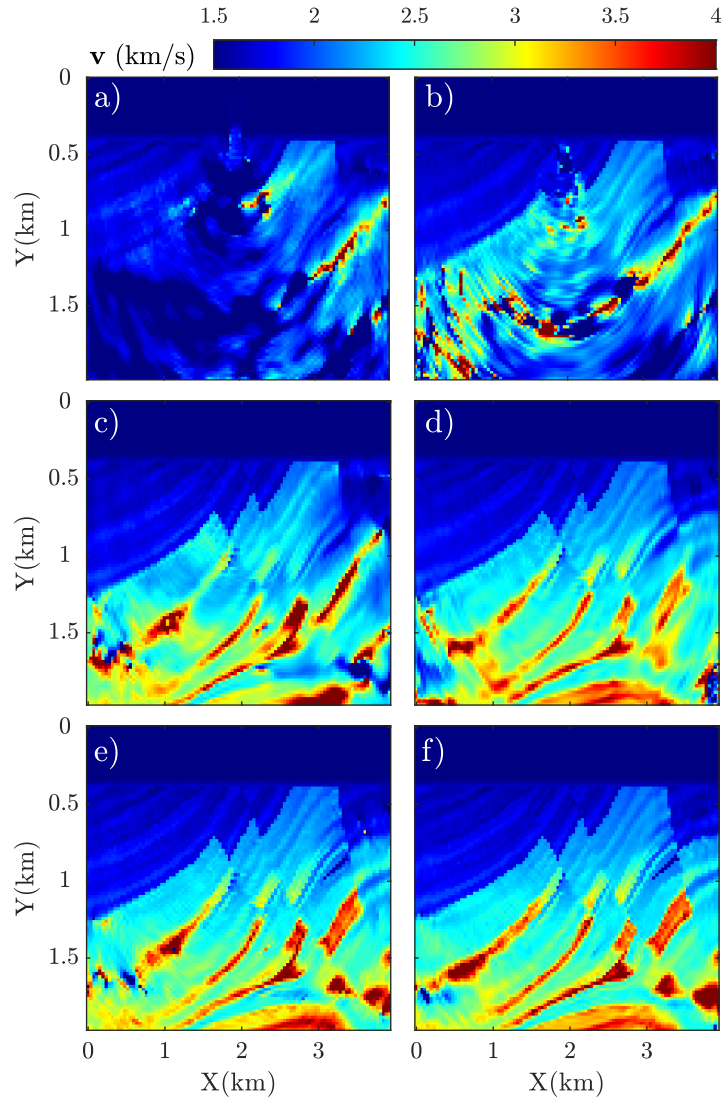
\plot{fig8}{width=0.62\textwidth}{Recovered velocity model computed with the reconstructed wavefields by diiferent approaches. (a) PINN with Q = $\infty$, (b) PINN with Q = 15, (c) PINN + PML with Q = $\infty$, (d) PINN + PML with Q = 15, (e) PINN + PML + quadratic NN with Q = $\infty$, (f) PINN + PML + quadratic NN with Q = 15.}
%
%
%
%
Without PML conditions and quadratic NN, the PINN wavefield can hardly recover the correct velocity model no matter in acoustic (Figures~\ref{fig:fig8}a) or visco-acoustic (Figures~\ref{fig:fig8}b) medium. Adding PML significantly improves the velocity reconstruction, but the velocity contrasts are blurred in acoustic (Figures~\ref{fig:fig8}c) and visco-acoustic (Figures~\ref{fig:fig8}d) medium. Using jointly quadratic NN and PML further improves the velocity reconstruction, in particular in the deep part at the reservoir level in acoustic (Figures~\ref{fig:fig8}e) and visco-acoustic (Figures~\ref{fig:fig8}f) medium, despite some artifacts in the shallow region remaining in Figure~\ref{fig:fig8}e. More insights about how quadratic NN improves the results of PINN are illustrated in the final discussion section. Compared to the velocity models of Figures~\ref{fig:fig8}c and \ref{fig:fig8}e (simulation without attenuation), those of Figures~\ref{fig:fig8}d and \ref{fig:fig8}f (simulation with attenuation) show an improvement in the most contrasted area of the model and near the surface. Compared to Figure~\ref{fig:fig8}e, the artifacts in the shallow region are also mitigated in Figure~\ref{fig:fig8}f.
This further illustrates how the reconstructed wavefields are improved by PML, quadratic NN and attenuation effects. 

Figure~\ref{fig:fig9} shows the evolution of the loss functions with iterations for the above six cases. Although the PML condition can improve the performance of PINN, it somehow causes a slow convergence rate in both the attenuation and non-attenuation cases,  especially during the early iterations. Combining PML with quadratic NN can significantly accelerate the convergence rate.
%
%
%
%
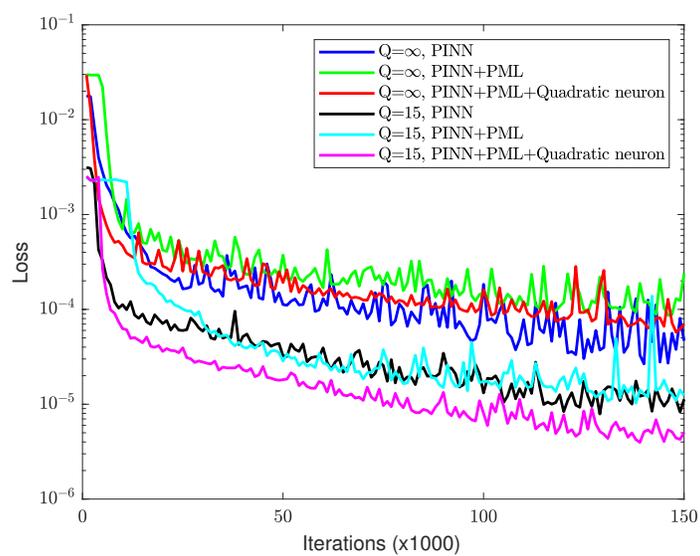
\plot{fig9}{width=0.6\textwidth}{The PINNs losses of different approaches during training.}

\section{Discussion}
\subsection{How do non-smooth velocities affect the results?}
To suggest an answer to the above question, let's come back to equation \ref{1}. In the PINN method, $\delta{u}$ is the output of the NN and the solution we are looking for. All the other parameters are known. Therefore, the left-hand side term of equation \ref{1} contains network parameters, unlike the right-hand side term. Let us say that the left-hand side term represents some operations on NN (multiplication and derivation) and the right-hand side term represents the \textit{label} of the NN after operations by some abuse of language. It should be noted that both the right- and left-hand side terms contain the model $m$. Considering the poor performance of PINN in non-smooth $m$, we should wonder which of the two terms (operations versus labels) drives PINN to a wrong solution? 
To answer this question, we consider the following two equations: 
\begin{equation}\label{12}
\begin{aligned}
\left(\omega^{2} m[\mathbf{x}] +\nabla^{2}\right) \delta {u}[\mathbf{x}] =-\omega^{2} \left(m_{s}[\mathbf{x}]-m_{0}[\mathbf{x}]\right) u_{0}[\mathbf{x}],
\end{aligned}
\end{equation}
\begin{equation}\label{13}
\begin{aligned}
\left(\omega^{2} m_{s}[\mathbf{x}] +\nabla^{2}\right) \delta {u}[\mathbf{x}] =-\omega^{2} \left(m[\mathbf{x}]-m_{0}[\mathbf{x}]\right) u_{0}[\mathbf{x}],
\end{aligned}
\end{equation}
\noindent where $m$ is non-smooth Marmousi model as shown in Figure~\ref{fig:fig3}a  and $m_{s}$ is smooth Marmousi model as shown in Figure~\ref{fig:fig2}a. Figures~\ref{fig:fig10}c and \ref{fig:fig10}d show the estimated solutions of equations \ref{12} and \ref{13} by PINN without using PML conditions or quadratic neurons. Apart from the velocity model, all the parameters are the same as those used in the example of Figure~\ref{fig:fig2}. The left-hand side term in equations \ref{12} and \ref{13} contains the direct operation on the output of NN, and the right-hand side terms do not change during training, like a label.
Based on the universal approximation theorem, if the label or the objective solution is not smooth enough along the input space, we can improve the performance of NN by increasing its size.
However, Figure~\ref{fig:fig10} shows that it is the non-smooth velocity in the \textit{operation} term instead of the \textit{label} term that causes the poor performance of PINN. 
This may explain why we can not improve the PINN results by simply increasing the size of the NN. Further research about the \textit{operation} term in PINN will be necessary during future work.
%
%
%
\plot{fig10}{width=0.62\textwidth}{Real parts of solutions from equations \ref{12} and \ref{13}. (a) FDFD solution of equations \ref{12},  (b) FDFD solution of equations \ref{13}, (c) PINN solution of equations \ref{12}, (d) PINN solution of equations \ref{13}, (e) Difference between Figures \ref{fig:fig10}a and \ref{fig:fig10}c, (f) Difference between Figures \ref{fig:fig10}b and \ref{fig:fig10}d.}
%
%
%

\subsection{How do PMLs improve the accuracy of PINN?}
Let's consider the total-field wave equation:
\begin{equation}\label{14}
\begin{aligned}
\left(\omega^{2} m[\mathbf{x}] +\nabla^{2}\right){u}[\omega,\mathbf{x}] =b[\omega,\mathbf{x}].
\end{aligned}
\end{equation}
where $u$ is the full wavefield and $b$ is the physical source. To mitigate the singularity of the point source $b$, we implement the real part of $b$ with a two-dimensional Gaussian function and the imaginary part is set to zero. The velocity model is homogeneous and $m=1/(1.5)^2 s^2/km^2$ and the frequency is set to $3 Hz$. The training samples are the points on 100 by 100 mesh with a  grid interval of $h=20 m$. The NN is trained with 50000 epochs. Figures~\ref{fig:fig11} show the estimated solutions by PINN with and without PML conditions. Without PML conditions, the real part of the wavefield is inaccurately reconstructed while the imaginary part remains close to zero. As it is mentioned before, when solving the frequency-domain wave equation or the scattered-field equation with PINN, the real and imaginary parts of the wavefield are calculated separately when PML conditions are not used. Thus, the connection between the real and imaginary part of the wavefield is not \textit{informed} in the loss function of PINN. In other words, the physical constraints are not enough to find the correct solutions. After adding PMLs, the PINN method can successfully reconstruct the solution of the wave equation. Figure~\ref{fig:fig12a} shows the convergence procedure during training and the role of the PMLs. The NN first reconstructs the real part of the wavefield approximately, proceeding from the source to the boundaries while the imaginary part remains close to zero. Once the real part of the wavefield reaches the boundary, the NN starts updating the imaginary part of the wavefield from the boundary to the source through the constraint provided by the PML conditions while refining the real part of the wavefield accordingly.
%
%
%
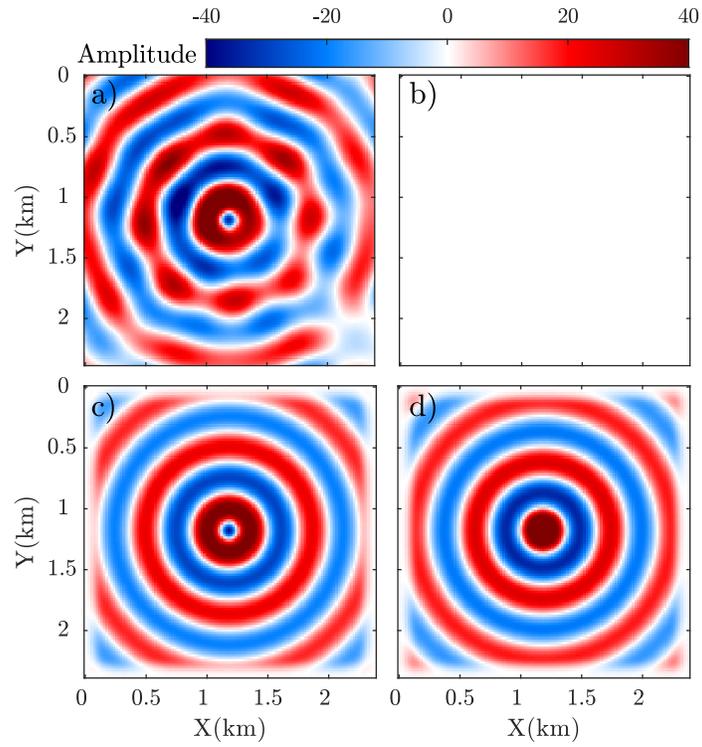
\plot{fig11}{width=0.6\textwidth}{Solutions of full wave equation by PINN and PINN + PML. (a) Real part by PINN, (b) imaginary part by PINN, (c) real part by PINN + PML, (d) imaginary part by PINN + PML.}
%
%
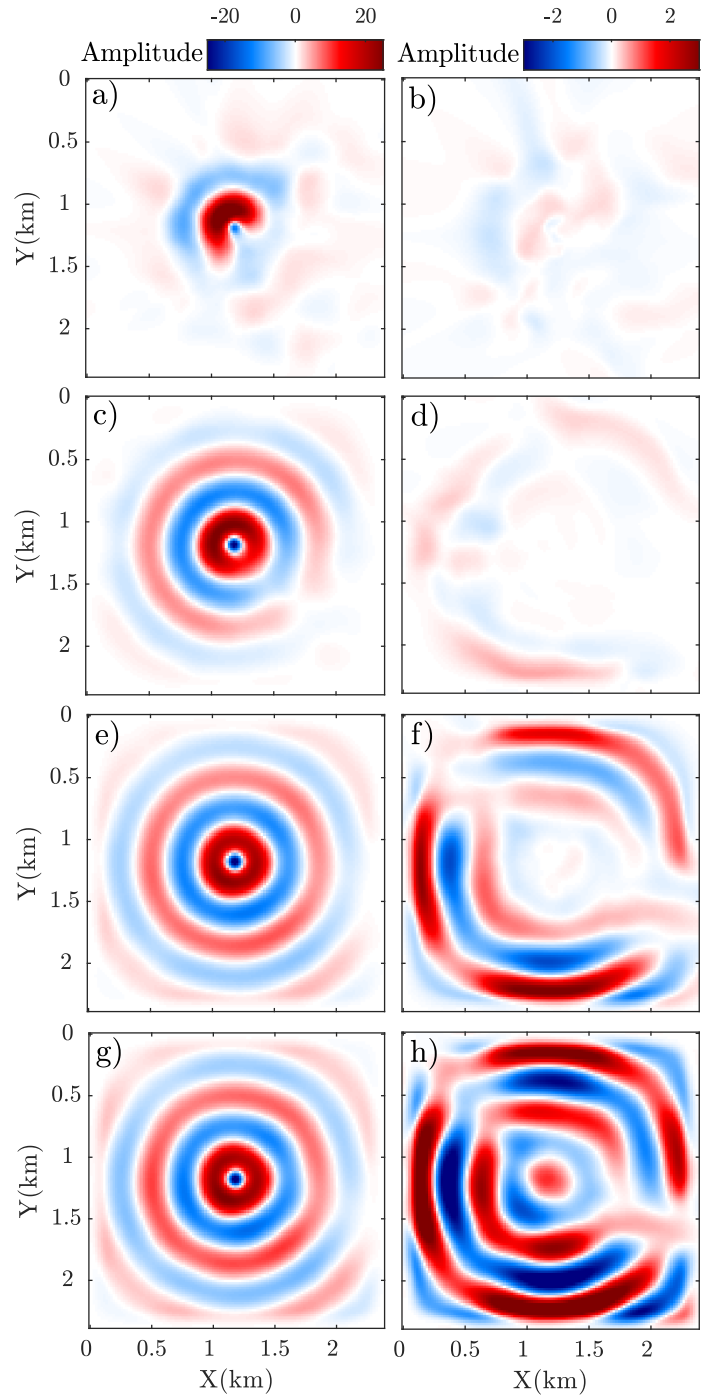
\plot{fig12a}{width=0.6\textwidth}{The training process during solving wave equation by PINN with PML. Left column (a,c,e,g) is real parts after 8000, 12000, 14000, and 20000 epochs. Right column (b,d,f,h) is imaginary parts after 8000, 12000, 14000, and 20000 epochs.}

\subsection{How does quadratic neuron improve the neural network?}
As mentioned before, the non-smooth velocity model causes a non-smooth (or discontinuous) Laplacian of wavefield (i.e., $\nabla^{2}\delta u$), which is difficult to approximate with PINN. There are two possible reasons: the network size is not large enough, or the network is not sensitive to the change of Laplacian during training. In our opinion, the network in this study has sufficient size. Instead, we believe that the quadratic NN improves the sensitivity of PINN to the change of Laplacian. 
To support this statement, we check in more detail the accuracy with which the Laplacian term is reconstructed by PINN.
Figures~\ref{fig:fig_vertical_logs}a and \ref{fig:fig_vertical_logs}b show the real part of the scattered wavefield and the Laplacian, respectively, in the non-attenuating Marmousi model along a vertical profile located at X=1200m. A direct comparison between the solutions computed with FDFD, PINN, PINN + PML and PINN + PML + quadratic NN is shown. Figures~\ref{fig:fig_vertical_logs}c and ~\ref{fig:fig_vertical_logs}d show the same information for the attenuating Marmousi model.
The wavefields from quadratic NN in both Figures~\ref{fig:fig_vertical_logs}a and \ref{fig:fig_vertical_logs}c have better results than those without using quadratic NN.
Compared to the real part of the scattered wavefields, non-smooth patterns are more obvious in the Laplacian of wavefields. This non-smooth behavior, as shown in Figures~\ref{fig:fig_vertical_logs}b and \ref{fig:fig_vertical_logs}d, is clearly better matched when quadratic NN is used.
Moreover, the computational overhead generated by the quadratic NN is negligible since we add only three weights in the neurons of the first layer.
%
%
%
%
%
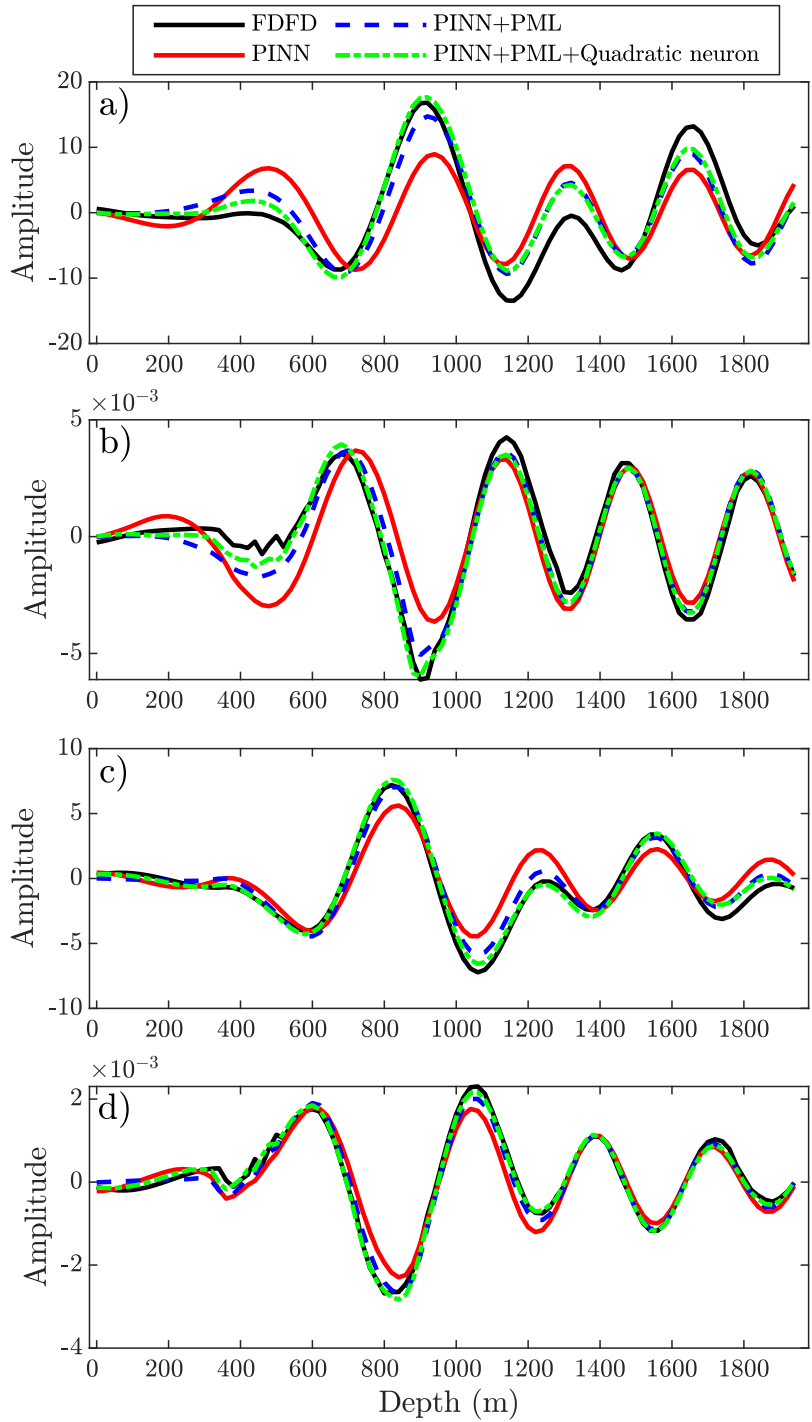
\plot{fig_vertical_logs}{width=0.7\textwidth}{The vertical profiles of real part and Laplacian of scattered wavefields in Marmousi model at X=1200m. (a) Real part with Q = $\infty$, (b) Laplacian with Q = $\infty$, (c) real part with Q = 15, (b) Laplacian with Q = 15, }

\subsection{Using pretraining for efficient wavefield simulation after model alteration}
Although PINN has shown its potential to solve the wave equation, there are still some remaining issues to solve. First, PINN recasts the forward problem as the minimization of a loss function for estimating the weights of the NN, which makes the computational cost to be a crucial issue. This issue can be greatly mitigated by pre-training. For example, we can firstly train a NN on a simple and smooth attenuating velocity model, e.g., a model where velocity linearly increases with depth as illustrated in Figure~\ref{fig:fig17}a. Due to the smoothness of the velocity model, the reconstructed wavefield by PINN + PML + quadratic NN (Figure~\ref{fig:fig17}d) matches reasonably well with the FDFD wavefield (Figure~\ref{fig:fig17}b) after 150000 epochs. Moreover, the differences between the two wavefields suggest that they mostly result from the different behavior of the PMLs in the FDFD and NN approaches, as shown in Figure~\ref{fig:fig17}e. The right colum It is worth mentioning that this process could be performed once and for all before performing a specific simulation. Then, we switch from the linear velocity model to a smooth version of the Marmousi model, as shown in Figure~\ref{fig:fig18}a, and train the pre-trained NN of the previous simulation (based on the linear velocity model) with the smooth Marmousi model. Figures~\ref{fig:fig18}d and \ref{fig:fig18}g show the reconstructed wavefield by PINN + PML + quadratic NN after 5000 and 20000 epochs, respectively. It takes 7 minutes for every 5000 epochs. A good agreement between the FDFD and the PINN wavefields is shown, although the number of epochs was reduced by around one order of magnitude compared to the first simulation. We continue this pre-training workflow by changing one more time velocity model by considering a sharper version of the Marmousi model as shown in Figure~\ref{fig:fig19}a.
We train the previous pre-trained NN of Figure~\ref{fig:fig18}d with 5000 epochs to reconstruct the wavefields of Figure~\ref{fig:fig19}d and train the NN of Figure~\ref{fig:fig18}g with 20000 epochs to reconstruct wavefields of Figure~\ref{fig:fig19}g.
Again, they (Figure~\ref{fig:fig19}d and \ref{fig:fig19}g) match fairly well with the FDFD wavefield (Figure~\ref{fig:fig19}b), although the limited number of epochs involved in the PINN simulation.
The right columns of Figure~\ref{fig:fig17}, \ref{fig:fig18} and \ref{fig:fig19} show the recovered velocity model from the lift wavefield of FDFD and PINNs. 
We conclude that we do not need to train a new NN when we perform a wavefield simulation after model alteration. Instead, we can train the NN that has been pre-trained during the simulation in the unaltered medium. In our case, this pre-training decreases the iteration count by around one order of magnitude and the more similar the models are, the less iterations the simulation takes. This pre-training strategy should find very interesting applications in FWI when the model perturbations become increasingly small as the FWI approaches the convergence point. Another application is time-lapse and target-oriented imaging, where the model alteration in the targeted area is generally very small.

%
%
%
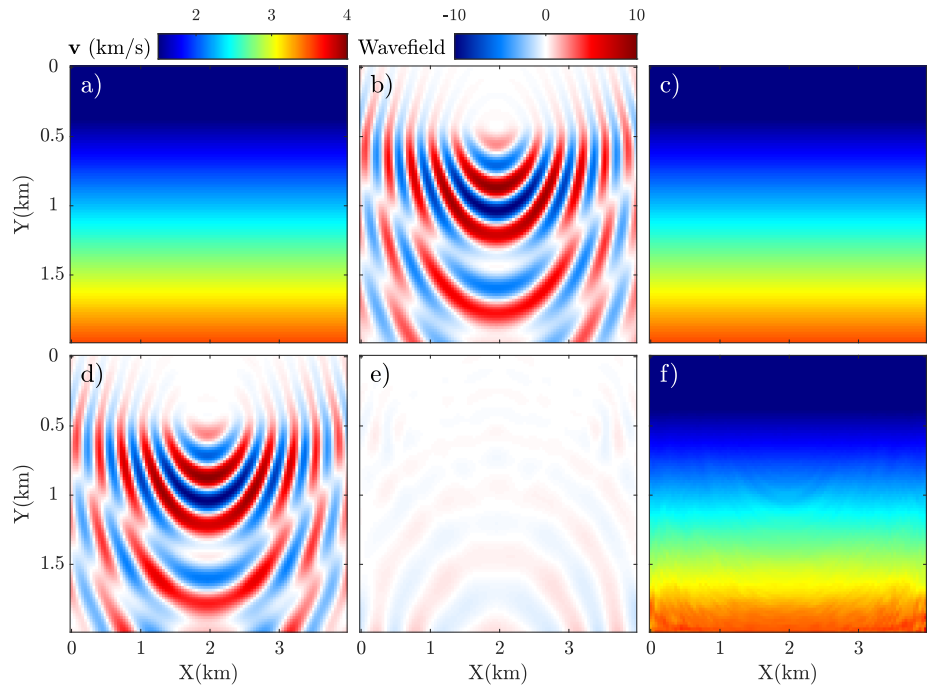
\plot{fig17}{width=0.8\textwidth}{The 1-D gradient velocity model and corresponding scattered wavefield with Q=15. (a) The true velocity model, (b) real part of wavefields by FDFD,  (c) recovered velocity models from Figure~\ref{fig:fig17}b, (d) real part of wavefields by PINN + PML + quadratic NN, (e) difference between Figures \ref{fig:fig17}b and \ref{fig:fig17}d, (f) recovered velocity models from Figure~\ref{fig:fig17}d. }
%
%
%
\plot{fig18}{width=0.8\textwidth}{The highly smoothed Marmousi model and corresponding scattered wavefield with Q=15. (a) The true velocity model, (b) real part of wavefield by FDFD,  (c) recovered velocity models from Figures \ref{fig:fig18}b, (d) real part of wavefield by PINN + PML + quadratic NN using 5000 iterations after Figure \ref{fig:fig17}d, (e) difference between Figures \ref{fig:fig18}b and \ref{fig:fig18}d, (f) the recovered velocity models from Figure~\ref{fig:fig18}d, (g-i) the same as (d-f), but using 20000 iterations after Figure \ref{fig:fig17}d.}
%
%
%
\plot{fig19}{width=0.8\textwidth}{The slightly smoothed Marmousi model and corresponding scattered wavefield with Q=15. (a) The true velocity model, (b) real part of wavefield by FDFD,  (c) recovered velocity models from Figures \ref{fig:fig19}b, (d) real part of wavefield by PINN + PML + quadratic NN using 5000 iterations after Figure \ref{fig:fig18}d, (e) difference between Figures \ref{fig:fig19}b and \ref{fig:fig19}d, (f) the recovered velocity models from Figure~\ref{fig:fig19}d, (g-i) the same as (d-f), but using 20000 iterations after Figure \ref{fig:fig18}g.}
%
%
%
%

Although PINN may show little advantage in solving simple forward problems (e.g., 2D, single source and single frequency), PINN may be beneficial for solving large-scale 3D forward problems involving multiple sources and frequencies. In this framework, the pre-training strategy introduced above may find applications to efficiently process multiple frequencies and sources in FWI applications. This will be investigated in future studies.

We have shown that the nonlinearity of Laplacian of the NN outputs with respect to the inputs has a significant effect on the performance of PINN. Since the quadratic neurons can improve the accuracy and nonlinearity of the Laplacian, a further study for designing a better NN architecture based on the characteristic of the Laplacian or derivative should be another interesting perspective.

\section{Conclusion}
We solve the frequency-domain scattered-field wave equation for non-smooth velocity models with PINN. We show numerically that the poor performance in contrasted media mainly results from the lack of physical constraint tying together the real and imaginary parts of the wavefields rather than from the size of the NN. This statement was further validated by tracking the history of reconstructing the real and imaginary parts of the wavefield over epochs. We mitigate this issue by explicitly implementing PML conditions in the loss function of PINN. We also use quadratic neurons in the first layer of the NN rather than linear neurons to simulate more efficiently the discontinuities in the Laplacian of the scattered wavefields. Adding physical attenuation is another leverage to couple the real and imaginary parts of the wavefields everywhere in the medium through complex wave speeds.
We show that pre-training can significantly mitigate the computational cost of PINN after model alteration. This opens the door to FWI and target-oriented imaging applications where the cost of the forward problem can be decreased as the inversion approaches the convergence point.
Although the accuracy of the wavefields computed with PINN does not reach yet that of classical numerical methods, it may be sufficient to start investigating FWI applications with PINN.

\section{ACKNOWLEDGMENTS}
We thank Arash Rezaei who helps to transfering the Matlab codes of FDFD to  Python. This work was (partially) supported by Beijing Natural Science Foundation No. Z210001, the WIND consortium (\hyperlink{https://www.geoazur.fr/WIND}{www.geoazur.fr/WIND}), and the China Scholarship Council. This work was supported by the French government, through the UCAJEDI Investments in the Future project  managed by the National Research Agency (ANR) under reference number ANR-15-IDEX-01. The authors are grateful to the OPAL infrastructure from Université Côte d’Azur and the Université Côte d’Azur’s Center for High-Performance Computing for providing resources and support.

\newpage
\bibliographystyle{seg} 
\bibliography{windbiblio,windtemp}

\newcommand{\SortNoop}[1]{}
\begin{thebibliography}{}
\itemsep0pt

\bibitem[Aghamiry et~al., 2019]{Aghamiry_2019_IWR}
Aghamiry, H., A. Gholami, and S. Operto,  2019, Improving full-waveform
  inversion by wavefield reconstruction with alternating direction method of
  multipliers: Geophysics, {\bfseries 84(1)}, R139--R162.

\bibitem[Alkhalifah et~al., 2021]{Alkhalifah_2021_WSM}
Alkhalifah, T., C. Song, U. bin Waheed, and Q. Hao,  2021, Wavefield solutions
  from machine learned functions constrained by the helmholtz equation:
  Artificial Intelligence in Geosciences, {\bfseries 2}, 11--19.

\bibitem[Amestoy et~al., 2016]{Amestoy_2016_FFF}
Amestoy, P., R. Brossier, A. Buttari, J.-Y. L'Excellent, T. Mary, L.
  M\'etivier, A. Miniussi, and S. Operto,  2016, Fast {3D} frequency-domain
  {FWI} with a parallel {B}lock {L}ow-{R}ank multifrontal direct solver:
  application to {OBC} data from the {N}orth {S}ea: Geophysics, {\bfseries 81},
  R363 -- R383.

\bibitem[Barron, 1993]{Barron_1993_UAB}
Barron, A.~R.,  1993, Universal approximation bounds for superpositions of a
  sigmoidal function: IEEE Transactions on Information Theory, {\bfseries 39},
  930--945.

\bibitem[B\'erenger, 1994]{Berenger_1994_PML}
B\'erenger, J.-P.,  1994, A perfectly matched layer for absorption of
  electromagnetic waves: Journal of Computational Physics, {\bfseries 114},
  185--200.

\bibitem[Berg and Nystr{\"o}m, 2019]{Berg_2019_DDP}
Berg, J., and K. Nystr{\"o}m,  2019, Data-driven discovery of pdes in complex
  datasets: Journal of Computational Physics, {\bfseries 384}, 239--252.

\bibitem[Burger et~al., 2021]{Burger_2021_CBD}
Burger, M., L. Ruthotto, S. Osher, et~al.,  2021, Connections between deep
  learning and partial differential equations: European Journal of Applied
  Mathematics, {\bfseries 32}, 395--396.

\bibitem[Cai et~al., 2022]{Cai_2022_PNN}
Cai, S., Z. Mao, Z. Wang, M. Yin, and G.~E. Karniadakis,  2022,
  Physics-informed neural networks (pinns) for fluid mechanics: A review: Acta
  Mechanica Sinica,  1--12.

\bibitem[Carcione, 2015]{Carcione_2015_WFR}
Carcione, J.~M.,  2015, Wave fields in real media, wave propagation in
  anisotropic, anelastic, porous and electromagnetic media, third edition ed.:
  Elsevier.

\bibitem[Chen et~al., 2013]{Chen_2013_OFD}
Chen, Z., D. Cheng, W. Feng, and T. Wu,  2013, An optimal 9-point finite
  difference scheme for the {H}elmholtz equation with {PML}: International
  Journal of Numerical Analysis \& Modeling, {\bfseries 10}.

\bibitem[Dolean et~al., 2015]{Dolean_2015_IDD}
Dolean, V., P. Jolivet, and F. Nataf,  2015, An introduction to domain
  decomposition methods - algorithms, theory, and parallel implementation:
  {SIAM}.

\bibitem[Erlangga and Herrmann, 2008]{Erlangga_2008_IMM}
Erlangga, Y.~A., and F.~J. Herrmann,  2008, An iterative multilevel method for
  computing wavefields in frequency-domain seismic inversion: SEG Technical
  Program Expanded Abstracts, {\bfseries 27}, 1956--1960.

\bibitem[Ernst and Gander, 2011]{Ernst_2011_Helm}
Ernst, O.~G., and M.~J. Gander,  2011, Why it is difficult to solve helmholtz
  problems with classical iterative methods: volume~{\bfseries 83} {\itshape
  of} Numerical Analysis of Multiscale Problems.

\bibitem[Fang and Zhan, 2019]{Fang_2019_PNN}
Fang, Z., and J. Zhan,  2019, A physics-informed neural network framework for
  pdes on 3d surfaces: time independent problems: IEEE Access, {\bfseries 8},
  26328--26335.

\bibitem[Gao et~al., 2021]{Gao_2021_PPG}
Gao, H., L. Sun, and J.-X. Wang,  2021, Phygeonet: physics-informed
  geometry-adaptive convolutional neural networks for solving parameterized
  steady-state pdes on irregular domain: Journal of Computational Physics,
  {\bfseries 428}, 110079.

\bibitem[Han et~al., 2018]{Han_2018_SHP}
Han, J., A. Jentzen, and W. E,  2018, Solving high-dimensional partial
  differential equations using deep learning: Proceedings of the National
  Academy of Sciences, {\bfseries 115}, 8505--8510.

\bibitem[Huang et~al., 2021]{Huang_2021_SPD}
Huang, X., H. Liu, B. Shi, Z. Wang, K. Yang, Y. Li, B. Weng, M. Wang, H. Chu,
  J. Zhou, et~al.,  2021, Solving partial differential equations with point
  source based on physics-informed neural networks: arXiv preprint
  arXiv:2111.01394.

\bibitem[Jagtap et~al., 2020]{Jagtap_2020_AAF}
Jagtap, A.~D., K. Kawaguchi, and G.~E. Karniadakis,  2020, Adaptive activation
  functions accelerate convergence in deep and physics-informed neural
  networks: Journal of Computational Physics, {\bfseries 404}, 109136.

\bibitem[Karimpouli and Tahmasebi, 2020]{Karimpouli_2020_PIM}
Karimpouli, S., and P. Tahmasebi,  2020, Physics informed machine learning:
  Seismic wave equation: Geoscience Frontiers, {\bfseries 11}, 1993--2001.

\bibitem[Karniadakis et~al., 2021]{Karniadakis_2021_PIM}
Karniadakis, G.~E., I.~G. Kevrekidis, L. Lu, P. Perdikaris, S. Wang, and L.
  Yang,  2021, Physics-informed machine learning: Nature Reviews Physics,
  {\bfseries 3}, 422--440.

\bibitem[Kharazmi et~al., 2021]{Kharazmi_2021_HVP}
Kharazmi, E., Z. Zhang, and G.~E. Karniadakis,  2021, hp-vpinns: Variational
  physics-informed neural networks with domain decomposition: Computer Methods
  in Applied Mechanics and Engineering, {\bfseries 374}, 113547.

\bibitem[Kostin et~al., 2019]{Kostin_2019_DFA}
Kostin, V., S. Solovyev, A. Bakulin, and M. Dmitriev,  2019, {Direct
  frequency-domain {3D} acoustic solver with intermediate data compression
  benchmarked against time-domain modeling for full-waveform inversion
  applications}: Geophysics, {\bfseries 84(4)}, T207--T219.

\bibitem[Le~Cun et~al., 2015]{Lecun_2015_DL}
Le~Cun, Y., Y. Bengio, and G. Hinton,  2015, Deep learning: Nature, {\bfseries
  521}, 436--444.

\bibitem[Long et~al., 2019]{Long_2019_PLP}
Long, Z., Y. Lu, and B. Dong,  2019, Pde-net 2.0: Learning pdes from data with
  a numeric-symbolic hybrid deep network: Journal of Computational Physics,
  {\bfseries 399}, 108925.

\bibitem[Long et~al., 2018]{Long_2018_PLP}
Long, Z., Y. Lu, X. Ma, and B. Dong,  2018, Pde-net: Learning pdes from data:
  International Conference on Machine Learning, PMLR, 3208--3216.

\bibitem[Mao et~al., 2020]{Mao_2020_PNN}
Mao, Z., A.~D. Jagtap, and G.~E. Karniadakis,  2020, Physics-informed neural
  networks for high-speed flows: Computer Methods in Applied Mechanics and
  Engineering, {\bfseries 360}, 112789.

\bibitem[Moseley et~al., 2020]{Moseley_2020_SWE}
Moseley, B., A. Markham, and T. Nissen-Meyer,  2020, Solving the wave equation
  with physics-informed deep learning: arXiv preprint arXiv:2006.11894.

\bibitem[Operto and Miniussi, 2018]{Operto_2018_MFF}
Operto, S., and A. Miniussi,  2018, On the role of density and attenuation in
  {3D} multi-parameter visco-acoustic {VTI} frequency-domain {FWI}: an {OBC}
  case study from the {North Sea}: Geophysical Journal International,
  {\bfseries 213}, 2037--2059.

\bibitem[Pang et~al., 2019]{Pang_2019_FFP}
Pang, G., L. Lu, and G.~E. Karniadakis,  2019, fpinns: Fractional
  physics-informed neural networks: SIAM Journal on Scientific Computing,
  {\bfseries 41}, A2603--A2626.

\bibitem[Plessix, 2007]{Plessix_2007_HIS}
Plessix, R.~E.,  2007, A {H}elmholtz iterative solver for {3D} seismic-imaging
  problems: Geophysics, {\bfseries 72}, SM185--SM194.

\bibitem[Plessix, 2017]{Plessix_2017_CAT}
--------, 2017, Some computational aspects of the time and frequency domain
  formulations of seismic waveform inversion, {\itshape in} Modern solvers for
  {H}elmholtz problems, Geosystems Mathematics: Springer,  159--187.

\bibitem[Raissi et~al., 2019]{Raissi_2019_PNN}
Raissi, M., P. Perdikaris, and G.~E. Karniadakis,  2019, Physics-informed
  neural networks: A deep learning framework for solving forward and inverse
  problems involving nonlinear partial differential equations: Journal of
  Computational Physics, {\bfseries 378}, 686--707.

\bibitem[Rasht-Behesht et~al., 2021]{Rasht_2021_PNN}
Rasht-Behesht, M., C. Huber, K. Shukla, and G.~E. Karniadakis,  2021,
  Physics-informed neural networks (pinns) for wave propagation and full
  waveform inversions: arXiv preprint arXiv:2108.12035.

\bibitem[Rasht-Behesht et~al., 2022]{Rasht_2022_PNN}
--------, 2022, Physics-informed neural networks (pinns) for wave propagation
  and full waveform inversions: Journal of Geophysical Research: Solid Earth,
  {\bfseries 127}, e2021JB023120.

\bibitem[Sirignano and Spiliopoulos, 2018]{Sirignano_2018_DGM}
Sirignano, J., and K. Spiliopoulos,  2018, Dgm: A deep learning algorithm for
  solving partial differential equations: Journal of Computational Physics,
  {\bfseries 375}, 1339--1364.

\bibitem[Song et~al., 2021]{Song_2021_SFA}
Song, C., T. Alkhalifah, and U.~B. Waheed,  2021, Solving the frequency-domain
  acoustic vti wave equation using physics-informed neural networks:
  Geophysical Journal International, {\bfseries 225}, 846--859.

\bibitem[Song and Alkhalifah, 2022]{Song_2022_WRI}
Song, C., and T.~A. Alkhalifah,  2022, Wavefield reconstruction inversion via
  physics-informed neural networks: IEEE Transactions on Geoscience and Remonte
  Sensing, {\bfseries 60}.

\bibitem[Sun et~al., 2020]{Sun_2020_SMF}
Sun, L., H. Gao, S. Pan, and J.-X. Wang,  2020, Surrogate modeling for fluid
  flows based on physics-constrained deep learning without simulation data:
  Computer Methods in Applied Mechanics and Engineering, {\bfseries 361},
  112732.

\bibitem[Tromp, 2019]{Tromp_2019_SWI}
Tromp, J.,  2019, {Seismic wavefield imaging of Earth's interior across
  scales}: Nature Reviews Earth \& Environment, {\bfseries 1}, 40--53.

\bibitem[Virieux et~al., 2017]{Virieux_2017_FWI}
Virieux, J., A. Asnaashari, R. Brossier, L. M\'etivier, A. Ribodetti, and W.
  Zhou,  2017, An introduction to {F}ull {W}aveform {I}nversion, {\itshape in}
  Encyclopedia of Exploration Geophysics: Society of Exploration Geophysics,
  R1--1--R1--40.

\bibitem[Virieux et~al., 2011]{Virieux_2011_RSP}
Virieux, J., H. Calandra, and R.~E. Plessix,  2011, A review of the spectral,
  pseudo-spectral, finite-difference and finite-element modelling techniques
  for geophysical imaging: Geophysical Prospecting, {\bfseries 59}, 794--813.

\bibitem[Virieux et~al., 2009]{Virieux_2009_TLE}
Virieux, J., S. Operto, H. {Ben Hadj Ali}, R. Brossier, V. Etienne, F.
  Sourbier, L. Giraud, and A. Haidar,  2009, Seismic wave modeling for seismic
  imaging: The Leading Edge, {\bfseries 28}, 538--544.

\bibitem[Voytan and Sen, 2020]{Voytan_2020_WPP}
Voytan, D., and M.~K. Sen,  2020, Wave propagation with physics informed neural
  networks: Presented at the SEG International Exposition and Annual Meeting,
  OnePetro.

\bibitem[Xu et~al., 2019]{Xu_2019_PIN}
Xu, Y., J. Li, and X. Chen,  2019, Physics informed neural networks for
  velocity inversion: Presented at the SEG International Exposition and Annual
  Meeting, OnePetro.

\bibitem[Zeng et~al., 2001]{Zeng_2001_APM}
Zeng, Y., J. He, and Q. Liu,  2001, The application of the perfectly matched
  layer in numerical modeling of wave propagation in poroelastic media:
  Geophysics, {\bfseries 66}, 1258--1266.

\bibitem[Zhang et~al., 2020]{Zhang_2020_LMS}
Zhang, D., L. Guo, and G.~E. Karniadakis,  2020, Learning in modal space:
  Solving time-dependent stochastic pdes using physics-informed neural
  networks: SIAM Journal on Scientific Computing, {\bfseries 42}, A639--A665.

\bibitem[Zhu et~al., 2019]{Zhu_2019_PDL}
Zhu, Y., N. Zabaras, P.-S. Koutsourelakis, and P. Perdikaris,  2019,
  Physics-constrained deep learning for high-dimensional surrogate modeling and
  uncertainty quantification without labeled data: Journal of Computational
  Physics, {\bfseries 394}, 56--81.

\end{thebibliography}
\end{document}